\documentclass{iau}
\usepackage{graphicx}

\usepackage{natbib}
\title[Radio-Loud NLS1 Kinematics]{The Parsec-scale Structure, Kinematics, and Polarization of Radio-Loud Narrow-Line Seyfert~1 Galaxies}

\author[J.~L. Richards et al.]
{J.~L. Richards$^{\rm 1,a}$, M.~L.~Lister$^1$, T.~Savolainen$^2$, D.~C.~Homan$^3$, M.~Kadler$^4$, T.~Hovatta$^{5}$, A.~C.~S.~Readhead$^6$, T.~G.~Arshakian$^7$, and V.~Chavushyan$^8$}

\affiliation{
  $^1$Department of Physics and Astronomy, Purdue University, West Lafayette, Indiana, USA\\[\affilskip]
  $^2$Max-Planck-Institut f\"ur Radioastronomie, Bonn, Germany\\[\affilskip]
  $^3$Department of Physics, Denison University, Granville, Ohio, USA\\[\affilskip]
  $^4$Institut f\"ur Theoretische Physik und Astrophysik, Universit\"at W\"urzburg, W\"urzburg, Germany\\[\affilskip]
  $^5$Aalto University Metsah\"ovi Radio Observatory, Kylm\"al\"a, Finland\\[\affilskip]
  $^6$Department of Astronomy, California Institute of Technology, Pasadena, California, USA\\[\affilskip]
  $^7$I. Physikalisches Institut, Universit\"at zu K\"oln, K\"oln, Germany\\[\affilskip]
  $^8$Instituto Nacional de Astrof\'isica, \'Optica y Electr\'onica, Puebla, Mexico\\[\affilskip]
  $^a$email: {\tt jlr@purdue.edu}\\[\affilskip]
}

\pubyear{2014}
\volume{313}  
\pagerange{}
\jname{Extragalactic Jets from Every Angle}
\editors{F.~Massaro, C.~C. Cheung, E.~Lopez, \& A.~Siemiginowska, eds.}
\begin{document}

\maketitle

\begin{abstract}
Several narrow-line Seyfert 1 galaxies (NLS1s) have now been detected in gamma rays, providing firm evidence that at least some of this class of active galactic nuclei (AGN) produce relativistic jets. The presence of jets in NLS1s is surprising, as these sources are typified by comparatively small black hole masses and near- or super-Eddington accretion rates. This challenges the current understanding of the conditions necessary for jet production. Comparing the properties of the jets in NLS1s with those in more familiar jetted systems is thus essential to improve jet production models. We present early results from our campaign to monitor the kinematics and polarization of the parsec-scale jets in a sample of 15 NLS1s through multifrequency observations with the Very Long Baseline Array. These observations are complemented by fast-cadence 15~GHz monitoring with the Owens Valley Radio Observatory 40~m telescope and optical spectroscopic monitoring with with the 2~m class telescope at the Guillermo Haro Astrophysics Observatory in Cananea, Mexico.
\end{abstract}

\firstsection 
\section{Introduction}
The existence of jets in narrow-line Seyfert~1 galaxies (NLS1s)
presents a challenge to our understanding of the relativistic jet
phenomenon in active galactic nuclei (AGN). NLS1s are similar to
ordinary Seyfert~1 galaxies but have unusually narrow broad lines
($\mathrm{FWHM(H\beta)<2000~km\,s^{-1}}$). NLS1s have many properties
associated with weak or absent jets~\citep[e.g.,][]{lacy_radio_2001,
  veron-cetty_are_2001}: they lie at the low end of the AGN black hole
mass range~\citep[$10^{5}-10^{8}~M_{\odot}$;
e.g.,][]{zhou_comprehensive_2006}, accrete at a substantial fraction
of the Eddington limit~\citep[e.g.,][]{boller_soft_1996}, and are
likely found in spiral or disk host
galaxies~\citep{ohta_bar_2007}. However, about 7\% are radio
loud~\citep{komossa_radio-loud_2006} and among these, evidence of jets
in these sources has been found, including from the detection of GeV
gamma-rays from several radio-loud
NLS1s~\citep[e.g.][]{fermi_nls1_class_2009}. While a number of
radio-loud NLS1s have been found to show jet morphologies on parsec
scales~\citep[e.g.,][]{doi_japanese_2007, gu_chen_2010,
  doi_very_2011}, a comprehensive study of the properties of these
jets has not been carried out. To address this, we are undertaking an
observation campaign to characterize the parsec-scale radio jet
morphologies and kinematics in a sample of radio-loud NLS1s.

Our observation program centers on a roughly year-long monitoring
campaign with the Very Long Baseline Array (VLBA), augmented with
fast-cadence single dish monitoring with the Owens Valley Radio
Observatory 40~m telescope. Optical spectroscopic monitoring of
several sources with the 2~m-class telescope at the Guillermo Haro
Optical Observatory (GHAO) was planned, but only one
(\mbox{J0324+3410}) turned out to be bright enough. We do not discuss
those results in these proceedings. Incidental to the VLBA program, we
also made snapshot observations of three NLS1s with the Karl G. Jansky
Very Large Array (VLA).

The sample, listed in Table~\ref{tab:sample}, comprises the known
radio-loud NLS1s above the celestial equator with a high-frequency
(1.4--15~GHz) archival flux density brighter than 30~mJy. These range
in redshift from 0.061 to 0.799, reaching the highest redshifts at
which NLS1s can be identified. The sample includes all three
northern-sky gamma-ray NLS1s in the 2-year \emph{Fermi} Large Area
Telescope (LAT) AGN Catalog~\citep{2LAC} and several gamma-ray
detections reported by \citet{foschini_evidence_2011}.

\begin{table}
  \setlength{\tabcolsep}{10pt} 
  \caption{Radio-Loud NLS1 monitoring sample (J2000 names).\label{tab:sample}}
  \begin{center}
    \begin{tabular}{c c c c} \hline
      J0324+3410$^*$ & J0948+0022$^*$ & J1435+3131 & J1629+4007 \\
      J0814+5609 & J0953+2836 & J1443+4725 & J1644+2619 \\
      J0849+5108$^*$ & J1047+4725 & J1505+0326$^*$ & J1722+5654 \\
      J0902+0443 & J1246+0238$^*$ & J1548+3511 \\
      \hline
    \end{tabular}

    \footnotesize{$^*$ gamma-ray detection reported in \citet{2LAC} or \citet{foschini_evidence_2011}.}
  \end{center}
\end{table}

\section{Parsec-Scale and Single-Dish Monitoring}
%
%
The core of our program is the multifrequency, polarimetric monitoring
with the VLBA. We will observe a total of 7 epochs with roughly
2~month separation, which will provide a total span of about a
year. The first observations were in February 2014, and as of October
we have completed four epochs. In each 24~h session, we interleave
observations of our program sources and calibrators to provide broad
$(u,v)$ coverage of each source. Each scan includes observations at
1.3, 2, 4, and 6~cm, all in full polarization with 2-bit recording and
256~MHz bandwidth in each polarization. This uses the full 2~Gbps data
rate achievable with the VLBA digital back-end. Total intensity image
noise levels of 0.07--0.1~mJy\,beam$^{-1}$ are obtained in all
bands. Phase referencing using nearby bright, compact targets is
employed for the fainter sources in the program.

\begin{figure}
  \centering
  \includegraphics[height=3in]{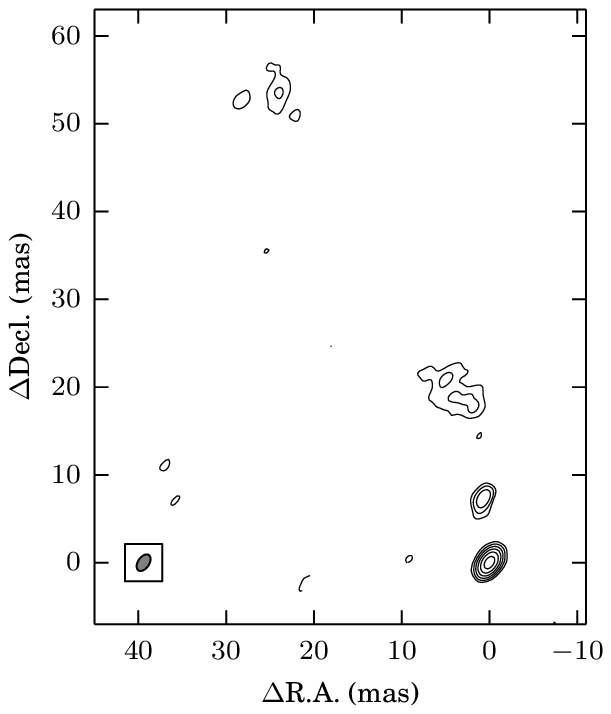}\hfill
  \includegraphics[height=3in]{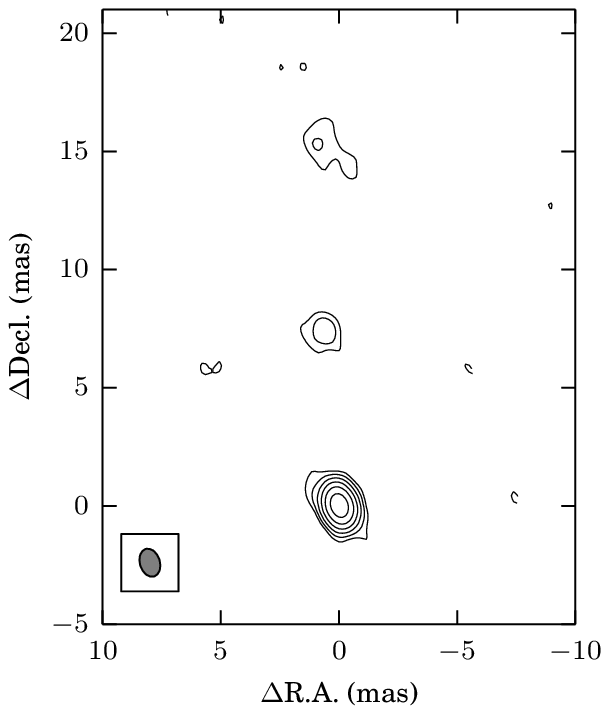}
  \caption{Preliminary VLBA total intensity maps of
    \mbox{J1548+3511}. Contours increase by factors of
    two. \textit{Left:} 4~cm map from the first epoch (8~February
    2014), showing emission extended to about
    60~arcsec. \textit{Right:} 2~cm map from the third epoch (30~June
    2014), showing good morphological agreement with the inner region
    of the 4~cm map. \label{fig:vlba_maps}}
  \label{fig:J1548+3511}
\end{figure}

Currently, data reduction for the completed VLBA epochs is under
way. Preliminary sample intensity maps are shown in
Figure~\ref{fig:vlba_maps}.  All of the sources have been detected,
and a variety of morphologies are found. Most resemble the compact,
core-dominated structures commonly found in
blazars~\citep[e.g.,][]{lister_MOJAVE_2005}. In all the resolved
cases, the morphology is consistent with a one-sided core-jet
structure. This resolved emission is found in 13 of the 15 sources,
making these promising candidates for our planned kinematics studies.

%
In addition to the VLBA monitoring, all the sources in our sample are
being monitored at 2~cm with the OVRO 40~m telescope as part of the
ongoing blazar monitoring program~\citep{richards_blazars_2011}. These
sources are observed about twice per week with a noise level of
3--5~mJy. We detect all our sources with mean flux densities between
11 and 480~mJy. Six of our 15 sources are too faint for variability
studies. In five of the nine brighter sources we detect variability
with amplitudes of 13--38\%, measured using the intrinsic modulation
index introduced in~\citet{richards_blazars_2011}. This measure
estimates the ratio of the standard deviation to the mean flux density
after removing the effect of observational uncertainties. We find
upper limits for the remaining four sources constraining them to be
less than 5--20\% variable. This range in variability amplitudes is
comparable to that seen in blazars in the OVRO 40~m program.

\section{Observations of Kiloparsec-Scale Extension}
At the outset of our program, extended radio emission on kiloparsec
scales had been reported for only six radio-loud
NLS1s~\citep{doi_kiloparsec-scale_2012}. We observed three sources from our
program for which the known radio positions were not precise enough
for correlation of our VLBA data (\mbox{J0953+2836},
\mbox{J1435+3131}, and \mbox{J1722+5654}). These observations were
carried out at 9~GHz while the VLA was being reconfigured from B
to A~array, its most extended and therefore highest resolution
configuration. The restoring beam for these observations was about
0.2~arcsec. This provided more than an order of magnitude increase in
resolution compared to most previous NLS1 observations. In 10~min
integrations with 2~GHz total bandwidth, we reached a noise level of
6--8~$\mu$Jy\,beam$^{-1}$ in the naturally weighted maps.

We detected significant kiloparsec-scale extended emission in all
three sources. All three showed core-dominated morphologies with
prominent two-sided edge-brightened lobes reminiscent of FR~II radio
galaxies. The full extents of the sources ranged from 4--10~arcsec,
corresponding to projected lengths of 23--70~kpc. Our observations,
which will be presented in detail in a forthcoming publication,
increase the number of radio-loud NLS1s with kiloparsec-scale
extensions by 50\%, from six to nine.

\section{Discussion}
The preliminary results of our observations provide further strong
evidence for the presence of relativistic jets in radio-loud
NLS1s. The one-sided morphologies we find on parsec-scales are
strongly reminiscent of those found in much larger, more powerful
blazars. These morphologies are consistent with the standard picture
of a pair of oppositely directed relativistic jets escaping from near
the central black hole. The morphologies on kiloparsec scales are also
consistent with this picture, showing that the parsec-scale jets
continue out to larger scales. Although we have only observed three
sources with the VLA, all three show kiloparsec-scale extension.  This
suggests that, in contrast to previous
findings~\citep{ulvestad_radio_1995}, when observed with the enhanced
sensitivity of the recently upgraded VLA, such extended emission is
common among the sources in our sample. In our OVRO single-dish
monitoring we find variability similar to that seen in many blazars,
furthering the comparison. It thus seems that although they are an
order of magnitude or more smaller than typical blazars and radio
galaxies, radio-loud NLS1s nonetheless frequently produce radio jets
that extend to kiloparsec scales. Our monitoring program will soon
enable a more detailed comparison of the jets in these classes of
sources, including their polarization and kinematic behavior.
\\

\noindent \textit{Acknowledgements:} This work was funded by the National
Aeronautics and Space Administration through \emph{Fermi} Guest
Investigator grant NNX13AO79G. The National Radio Astronomy
Observatory is a facility of the National Science Foundation operated
under cooperative agreement by Associated Universities, Inc. This
research has made use of the NASA/IPAC Extragalactic Database (NED)
which is operated by the Jet Propulsion Laboratory, California
Institute of Technology, under contract with the National Aeronautics
and Space Administration.

\bibliography{richards}
\end{document}